\documentclass[12pt,preprint]{aastex}

\slugcomment{Submitted to xx}
\shortauthors{Brinkworth et al.}
\shorttitle{The Dusty Disks Around Four SDSS White Dwarfs}

\begin{document}

\title{
A Spitzer Space Telescope study of the Debris Disks around four SDSS White Dwarfs 
}
 
\author{C. S. Brinkworth,\altaffilmark{1,2} 
B. T. G\"ansicke,\altaffilmark{3} J. M. Girven, \altaffilmark{1,3} D. W. Hoard, \altaffilmark{1} T. R. Marsh,\altaffilmark{3} S. G. Parsons \altaffilmark{3} \& D. Koester \altaffilmark{4} 
}
\altaffiltext{1}{Spitzer Science Center, California Institute of Technology,
Pasadena, CA 91125}
\altaffiltext{2}{NASA Exoplanet Science Institute, California Institute of Technology, Pasadena CA 91125}
\altaffiltext{3}{Dept of Physics and Astronomy, University of Warwick, Warwick,CV4 7AL,  United Kingdom}
\altaffiltext{4}{Institut f\"ur Theoretische Physik und Astrophysik, University of Kiel, 24098 Kiel, Germany}
\begin{abstract}

\end{abstract}

We present Spitzer Space Telescope data of four isolated white dwarfs that were previously known to harbor circumstellar gaseous disks. IRAC photometry shows a significant infrared excess in all of the systems, SDSS0738+1835, SDSS0845+2257, SDSS1043+0855 and SDSS1617+1620, indicative of a dusty extension to those disks. The 4.5$\mu$m excesses seen in SDSS0738, SDSS0845, and SDSS1617 are 7.5, 5.7 and 4.5 times the white dwarf contribution, respectively. In contrast, in SDSS1043, the measured flux density at 4.5 microns is only 1.7 times the white dwarf contribution. 

 We compare the measured IR excesses in the systems to models of geometrically thin, optically thick disks, and find that we are able to match the measured SEDs to within 3$\sigma$ of the uncertainties, although disks with unfeasibly hot inner dust temperatures generally provide a better fit than those below the dust sublimation temperature. Possible explanations for the dearth of dust around SDSS1043+0855 are briefly discussed. Including our previous study of SDSS1228+1040, all five white dwarfs with gaseous debris disks have significant amounts of dust around them. It is evident that gas and dust can coexist around these relatively warm, relatively young white dwarfs. 


\section{Introduction}

Since the discovery of the first dusty disk around the white dwarf G29-39 \citep{zuckerman87, graham90, telesco90, tokunaga90, reach05, vonhippel07}, dust disks have been detected around at least 26 isolated white dwarfs \citep{jura03, kilic05, becklin05, kilic06, jura07, vonhippel07, kilic07, farihi08, jura08, brinkworth09, farihi09, farihi10, melis10, vennes11, debes11a, xu11, kilic11, girven11}, offering clues to the fate of planetary systems when their host stars undergo post-main sequence evolution. The most likely origin for the disks is the tidal distruption of asteroids \citep{graham90, debes02, jura03}, and the subsequent accretion of this tidally distrupted material can explain the large metal abundances in the white dwarf atmospheres. 

The discovery of a gaseous disk around the hot white dwarf SDSS\,J1228+1040 \citep{gaensicke06} and our subsequent identification of a dusty extension to that disk \citep{brinkworth09} showed that debris disks are not restricted to low-temperature white dwarfs, as previously believed \citep{kilic06}, and dust and gas readily coexist in such a disk. Indeed, measured parameters for the 18 dusty white dwarfs \citep[see][for an overview]{farihi10} show that the presence of dust does not appear to correlate with temperature. 

Here we present infrared photometry and SED modelling for a further four white dwarfs similar to SDSS\,J1228+1040: SDSS\,J104341.53+085558.2, SDSS\,J084539.17+225728.0, \linebreak SDSS\,J161717.04+162022.4 and SDSS\,J073842.57+183509.6 (hereafter SDSS1043, SDSS0845, SDSS1617 and SDSS0738, respectively). As with SDSS\,J1228+1040, all have previously been identified from Sloan Digital Sky Survey spectroscopy, revealing Ca\,II 860\,nm emission lines from a gaseous disk component \citep{gaensicke07, gaensicke08, gaensicke12}, \citep{gaensicke11}, leading us to search for an infrared excess indicative of dusty extensions to these disks. SDSS0738 was independently discovered by \citet{dufour10} because of extremely strong metal pollution. \citet{dufour10} also reported an infrared excess in the near-IR, but did not notice the CaII emission of the gaseous disk component.  

 SDSS1228, SDSS1043 and SDSS1617 are all hydrogen-dominated DA white dwarfs, while SDSS0845 and SDSS0738 are helium-dominated DB white dwarfs. All are metal-polluted, all are moderately "warm", with temperatures ranging from 13,000 - 22,000\,K, and hence they are relatively young (cooling ages of $\sim$100-600\,Myr).

\section{Observations and Data Reduction}

\subsection{Spitzer Space Telescope}
 
We were awarded 2.7 hours in Cycle 4 (P40048), 0.7 hours in Cycle 5 (P50118) and 2.5 hours in Cycle 7 (P70012) to search for a dusty component to the disks around our four targets (see Table\,\ref{table:observations} for a full list of observations). For SDSS1043, we obtained Infrared Array Camera \citep[IRAC; see][]{fazio05} data in all four channels (3.6\,$\mu$m -- 8.0\,$\mu$m). For SDSS0845, we obtained IRAC data at 4.5\,$\mu$m and 8.0\,$\mu$m, Infrared Spectrograph \citep[IRS;][]{houck04} Blue Peak-Up Imaging at 16\,$\mu$m and Multiband Imaging Photometer \citep[MIPS;][]{rieke04} imaging at 24\,$\mu$m. For SDSS1617 and SDSS0738, we obtained Warm IRAC data at 3.6\,$\mu$m and 4.5\,$\mu$m. 

The IRAC data reduction for all sources was carried out on the individual corrected Basic Calibrated Data (CBCD; cbcd.fits) frames downloaded from the Spitzer archive. The CBCDs were initially combined with the standard SSC software, MOsaic and Point Source Extraction \citep[MOPEX;][]{makovoz06} with dual outlier rejection to create a single mosaicked image for each channel. This was used to determine the position of the target. All of the original downloaded CBCD frames were then corrected for array location dependence, using the correction frames supplied by the Spitzer Science Center (SSC). Aperture photometry was carried out on the corrected CBCD frames with IRAF, using an aperture radius of 3.0 pixels. The sky subtraction was applied using an annulus from 12-20 pixels in radius.

SDSS1043 was found to have a blended neighbour in the extraction aperture. We calculated the individual flux densities for each of the blended sources by measuring the combined flux, then splitting this by the ratio of peak brightness of a Gaussian fit to each source in each channel. 

The IRAC photometry for all sources was then converted from MJy sr$^{-1}$ to $\mu$Jy, and aperture-corrected using the standard aperture corrections provided by the SSC. We did not apply a pixel phase correction to the channel 1 data, but included an extra 1\% in our error budget to account for this \citep{reach05}. No colour correction was applied, since we quote the isophotal wavelengths, thus eliminating most of the colour dependency of the flux calibration and rendering the magnitude of the correction negligible compared to our uncertainties. Uncertainties were estimated using the RMS scatter in the photometry in the individual frames, divided by the square root of the number of frames. The quoted uncertainties are either our calculated uncertainty or the IRAC calibration uncertainty derived by \citet{reach05}, whichever was the larger. 

The IRS Peak-Up Imaging and MIPS data for SDSS0845 were combined in much the same way, using MOPEX to create a mosaic to measure the source positions, and carrying out the photometry on the individual Basic Calibrated Data (BCD) frames. The sky background was found to be variable across the IRS Peak-Up array, and so we created a sky flat by combining all of the dithers and rejecting the sources. This sky flat was subtracted from the individual frames before carrying out the photometry, bringing the sky background to zero. The photometry for both the IRS and MIPS data was obtained via IRAF, using an aperture radius of 3.0 pixels with no sky subtraction for IRS, and an aperture radius of 3$\arcsec$ (1.224 pixels) with a background aperture of 20$\arcsec$ -- 32$\arcsec$ for MIPS. After converting from MJy sr$^{-1}$ to $\mu$Jy, the flux densities were aperture-corrected to the calibration apertures using the standard corrections provided by the SSC. 

Finally, for all of the data for the four targets, we rejected the photometry from the original frames that had been flagged by the MOPEX Dual Outlier rejection algorithm. We take the unweighted mean of the remaining photometry in each channel as the quoted flux densities. An estimate of the uncertainties was measured using (RMS scatter) / sqrt(Number of frames), consistently giving uncertainties lower than the instrument calibration uncertainties. We therefore adopt the instrument calibration uncertainties from the SSC as the uncertainties on our data. For MIPS this is 4\%, and for IRSPUI this is 5\%. 

\subsection{AAT, SOFI, UKIDSS, GALEX, SDSS, 2MASS and WISE Data}

Near-infrared imaging of SDSS0845 was obtained at the Australian Astronomical Observatory's Anglo-Australian Telescope (AAT) using the InfraRed Imaging Spectrograph 2 (IRIS2) on 22 February, 2008, and retrieved from the public AAT archive. A 9-point dither pattern was used, with individual exposure times of 30s, 42s and 67s, and total exposure times of 270s, 378s and 603s in \textit{J}, \textit{H} and \textit{K$_{s}$}, respectively. SDSS1617 was observed in the K$_{s}$ band on 1 May, 2010 using SOFI on the ESO New Technology Telescope (NTT). A total of 20 5s images were obtained, offset randomly within a 15$\arcsec$ box.  The data were sky-median subtracted, astrometrically calibrated using the WCSTools\footnote{http://tdc-www.harvard.edu/wcstools} and co-added using SWarp\footnote{http://www.astromatic.net}. Object detection and the aperture photometry were done using the SExtractor \citep{bertin96}. Instrumental magnitudes were converted to the 2MASS system using a set of five bright stars near SDSS0845 and ten stars near SDSS1617. The uncertainty in the calibration is 0.02\,mag in all bands, and was added in quadrature to the statistical uncertainties in the photometry. The \textit{J}, \textit{H} and \textit{K$_{s}$} magnitudes, fluxes and uncertainties are reported in Table\,\ref{table:photometry}. 

Observations of SDSS1043 in $Y$, $J$, $H$ and $K$ bands have been obtained with UKIRT/WFCAM as part of UKIDSS \citep{lawrence07} on 22 May and 15 Dec, 2009 (Table\,\ref{table:photometry}). The GALEX \citep{martin05}, SDSS \citep{abazajian09} and 2MASS \citep{skrutskie06} archives and the WISE Preliminary Data Release were searched for additional photometry of our targets. The archival and WISE photometry are also listed in Table\,\ref{table:photometry}.

\subsection{Possible contamination of SDSS\,J0845+2257}\label{section:sdss0845contaminant}

The NIRI imaging of SDSS0845 obtained by \citet{melis10} revealed the presence of a faint galaxy at a $1.8\arcsec$ separation north-east of the white dwarf. We have downloaded the NIRI $J$, $H$, and $K_\mathrm{s}$-band images of SDSS0845 from the Gemini Science Archive, and, after correction for non-linearity using the python scripts provided by Gemini, processed them in the same manner as the AAT and NTT data described above. In order to measure the relative contributions of the star and the galaxy, first an annulus was defined, centred on the star, with inner and outer radii of 4 and 7" on the sky (large enough to avoid the galaxy) to measure the sky background. The star's contribution was then measured from the flux summed over a 1.6" diameter aperture, while the galaxy's flux was measured from the union of pixels covered by three overlapping circular apertures covering its major axis. To estimate the degree to which the galaxy flux was contaminated by light from the star, three identical control apertures were placed diametrically opposite to the galaxy apertures relative to the star. The instrumental magnitudes were photometrically calibrated using a bright 2MASS star in the field. We find $J=16.44\pm0.05$, $H=16.30\pm0.06$ and $K=16.04\pm0.07$ for SDSS0845, and $J=20.33\pm0.79$, $H=19.29\pm0.32$ and $K=18.65\pm0.24$ for the nearby galaxy. These results agree well with the NIRI magnitudes of \citet{melis10}, apart from the fact that we have somewhat larger uncertainties.

To estimate the contamination of the galaxy in the \textit{Spitzer} beam, we obtained the galaxy spectral energy distributions (SED) of Maraston (1998, 2005) based on the initial mass function of Kroupa (2001). We then redshifted all galaxy SEDs by redshifts $0\le z\le3$, in steps of 0.2, and discarded all combinations of redshift and age that would disagree with the observed $J-H$ and $H-K_\mathrm{s}$ colours and the apparent $K_\mathrm{s}$ magnitude. The NIRI data constrains the redshift of the galaxy to $1.4\la z\la3.0$, with the range $2.4-3.0$ only feasible for young ($<1$\,Gyr) galaxies. The spatial extension of the galaxy is suggestive of a redshift at the lower end of the range allowed by the NIRI photometry (Ferguson et al 2004). 

The contribution of the galaxy to the IRAC-1 and 2 channels is $0.04\pm0.01$\,mJy and $0.05\pm0.01$\,mJy, relatively independent of the galaxy type. The maximum contribution in the IRAC-3 and 4 channel is $\sim8$\,mJy for a young ($<1$\,Gyr) high-redshift ($z\sim3$) galaxy. However, given the combined constraint from morphology and near-infrared photometry, the contribution of the galaxy to the IRAC-3 and 4 channels is likely to be $<5$\,mJy. Galaxy models are poorly constrained in the mid-IR, and so we are unable to learn anything more abotu the galaxy from the IRS-PU and MIPS channels, other than the fact that the galaxy flux continues to drop out ot longer wavelengths. At these longer wavelengths, the SED is dominated by the flux density from the dust disk. In Figure\,\ref{figure:sdss0845} we show an 8 Gyr old, redshift z=1.6 galaxy as an example of the flux contribution to the Spitzer bands.

\section{Results and Modeling} \label{sec:models}

We plotted the SEDs of the four WDs, comparing them to models of isolated white dwarfs and other system components as described below. We define an infrared excess as significant if any of the infrared photometric points lies more than 3$\sigma$ above the model of the white dwarf. 

Where we detect an infrared excess, we attempt to model it using an optically thick, geometrically thin dusty disk, with a temperature profile $T_{disk} \propto r^{-\beta}$ where $\beta = 3/4$ \citep{jura03}, as previously described in \citet{brinkworth09}. The ratio of stellar radius to distance sets the absolute scale of the white dwarf photospheric flux, and is fixed in our modeling based on the best available parameters. The free parameters in the model are the inner disk radius ($R_{in}$), the outer disk radius ($R_{out}$), and the disk inclination ($i$). A grid of disk models was calculated with inner ($T_{in}$) and outer ($T_{out}$) disk temperatures, each corresponding to a value of $R_{in}$ and $R_{out}$ ranging from 100 to 1800\,K in steps of 50\,K, and inclination ranging from 0 to 90 degrees in steps of 5 degrees. $T_{out}$ was fixed to be cooler than $T_{in}$. A least $\chi^{2}$ method was used to fit the $H-, K-$ and IRAC-band infrared fluxes. See \citet{girven11} for further information on the fitting technique. 

The model is plotted as a dash-dot grey line in the figures.  For each of our white dwarf targets, we include two figures: one showing the best-fit model to the datapoints, regardless of physical feasibility, and one showing the closest possible physical model to that best fit. Analyses of the photospheric abundances of white dwarfs hosting dusty disks suggest that the bulk composition of the circumstellar material is dominated by Mg, O, Si, and Fe, similar to the bulk composition of the Earth \citep{zuckerman07, klein11}. The condensation temperature of the main minerals carrying these elements is ~1400K \citep{lodders03}, and we therefore do not expect that dusty debris discs around white dwarfs can survive temperatures much above that.  Hence, we define an inner dust disk temperature of 1400K as a physical limit. For each target, we also describe the coolest inner disk temperature that we were able to match to the data while staying consistent to within 3$\sigma$ of all the datapoints.

\subsection{SDSS\,J0738+1835}\label{section:sdss0738}

\begin{figure}
\plottwo{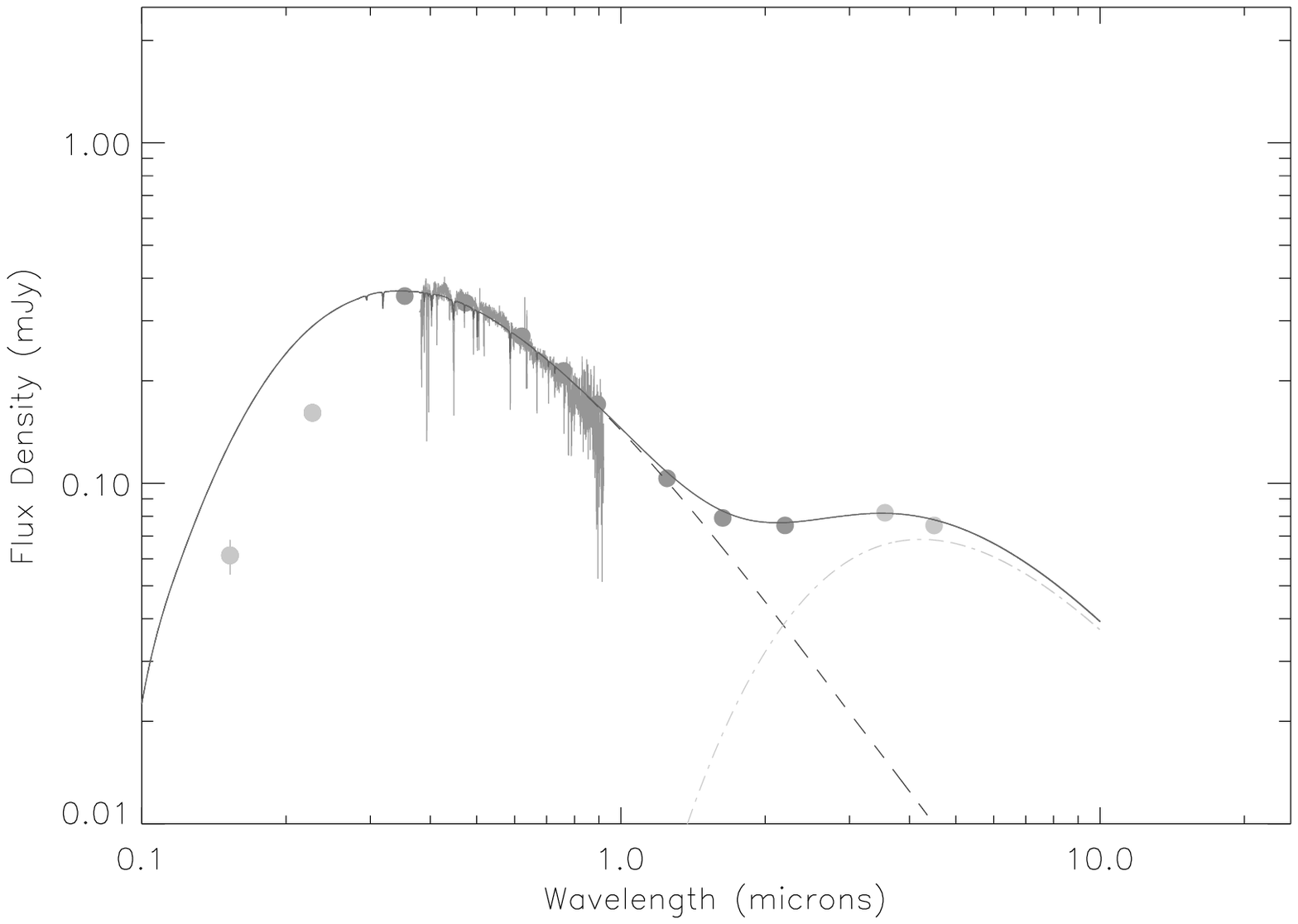}{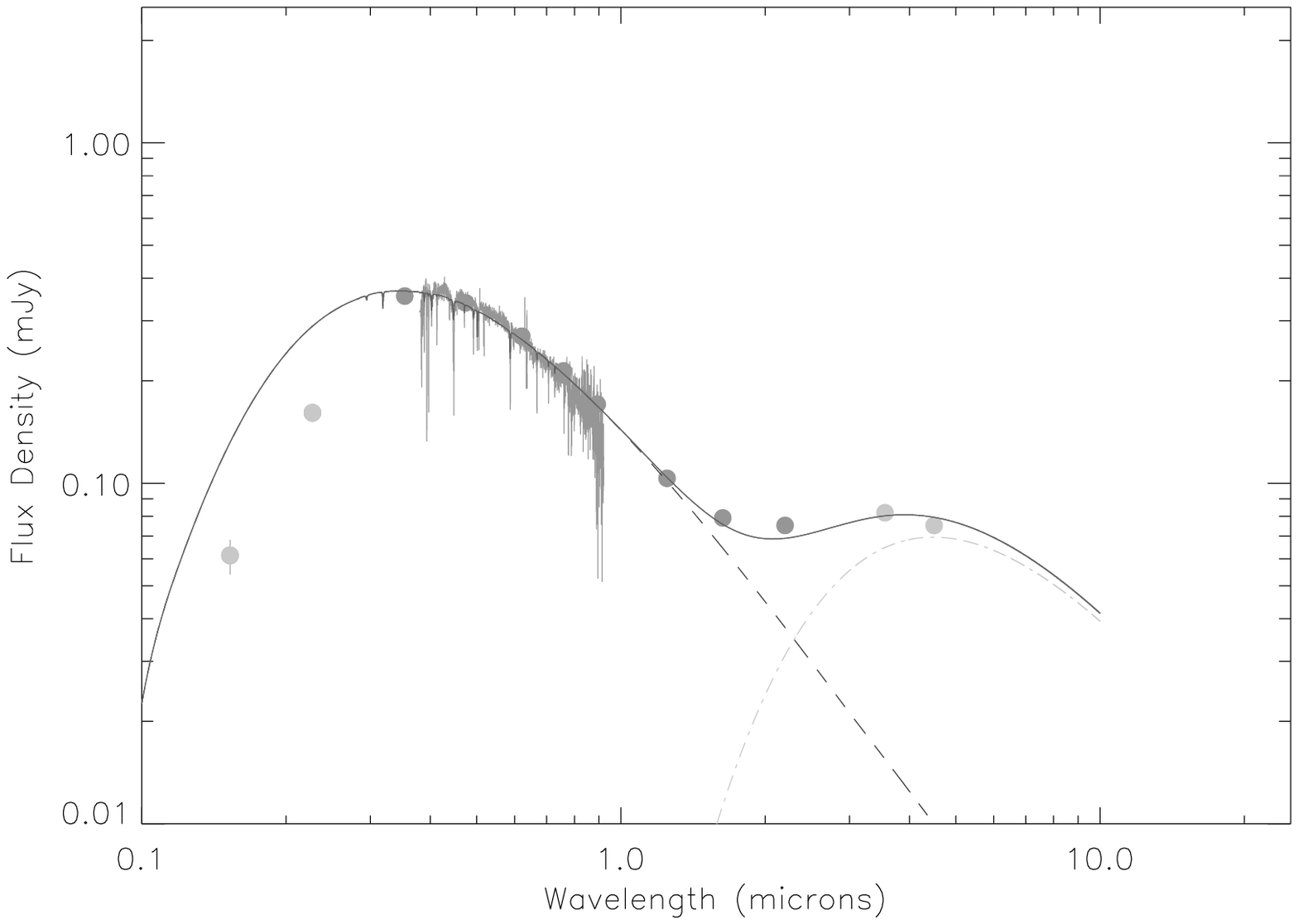}
\caption{The spectral energy distribution (SED) of SDSS\,J0738+1835 from the ultraviolet to the mid-infrared. See Table\,\ref{table:photometry} for the GALEX and \textit{Spitzer} data. The WD spectrum was taken from the SDSS Data Release 7 and is plotted along with the GALEX, \textit{Spitzer}, WISE and SDSS photometry, and the near-IR data from \citet{dufour10}. The SED shows a significant flux density excess above that expected from the white dwarf alone. The SED on the left is compared to a white dwarf + optically thick disk model with an inner dust temperature of 1800\,K, and outer disk temperature of 830\,K and an inclination of 58 degrees. The SED on the right limits the inner disk temperature to $T_{in}$ = 1400\,K, with $T_{out}$ = 930\,K and a face-on inclination of 0 degrees. See Section\,\ref{section:sdss0738} for the full range of possible models.  }
\label{figure:sdss0738}
\end{figure}

The flux densities from GALEX and Spitzer are given in Table\,\ref{table:photometry}, and are plotted in light grey in Figure\,\ref{figure:sdss0738}, along with the SDSS DR7 spectrum (grey line). We also included near-IR photometry from \citet{dufour10}, also plotted in dark grey. The DB white dwarf model is shown as a dashed black line, with T$_{eff}$ = 13600\,K, log(g) = 8.5, $R_{WD}$ = 0.00886\,R$_{\odot}$ \citep[Table 3; see][]{dufour10}. We use a distance of d = 120\,pc, rather than the 136\,pc$\pm22$ quoted in \citet{dufour10}, as this gives a better fit to the WD SED. This is still consistent with Dufour et al.'s value. 

The photometry of SDSS0738 shows a clear flux density excess in the infrared wavelengths, compared to the white dwarf model. The best fitting disk model (Figure\,\ref{figure:sdss0738}a) has $T_{in}$ = 1800\,K, $T_{out}$ = 830\,K and an inclination of 60 degrees, which is consistent with the modeling results from \citet{dufour10}. The best physically-motivated model with $T_{in}$ = 1400\,K, has $T_{out}$ = 930\,K and a face-on inclination of 0 degrees (Figure\,\ref{figure:sdss0738}b). This corresponds to inner and outer disk radii of $R_{in}$ = 12.4 $R_{WD}$ (0.11\,$R_\odot$) and $R_{out}$ = 21.3 $R_{WD}$ (0.19\,$R_{\odot}$). The coolest possible inner disk temperature is constrained by the need for a high enough flux density to match the JHK datapoints, and has $T_{in}$ = 1350\,K, $T_{out}$ = 880\,K and an inclination of 0 degrees. These models are both consistent with the modelling carried out by \citet{dufour12}, who used $T_{in}$ = 1600 $\pm$100\,K, $T_{out}$ = 900 +100/-200\,K and an inclination of 40$\pm$20 degrees.

Analysis of the CaII emission lines gives an outer radius of the gaseous disc component of $R_{out}\simeq 1R_{\odot} \sin^2 i$, i.e. a maximum value of  $R_{out}\sim110R_{WD}$,  \citep{gaensicke12}. While the CaII emission lines provide no formal lower limit on $R_{out}$, a low-inclination, practically face-on geometry seems unlikely, as it would imply a very narrow ring for the gaseous component, narrower than the dusty component. 

\subsection{SDSS\,J0845+2257}\label{section:sdss0845}

\begin{figure}
\plottwo{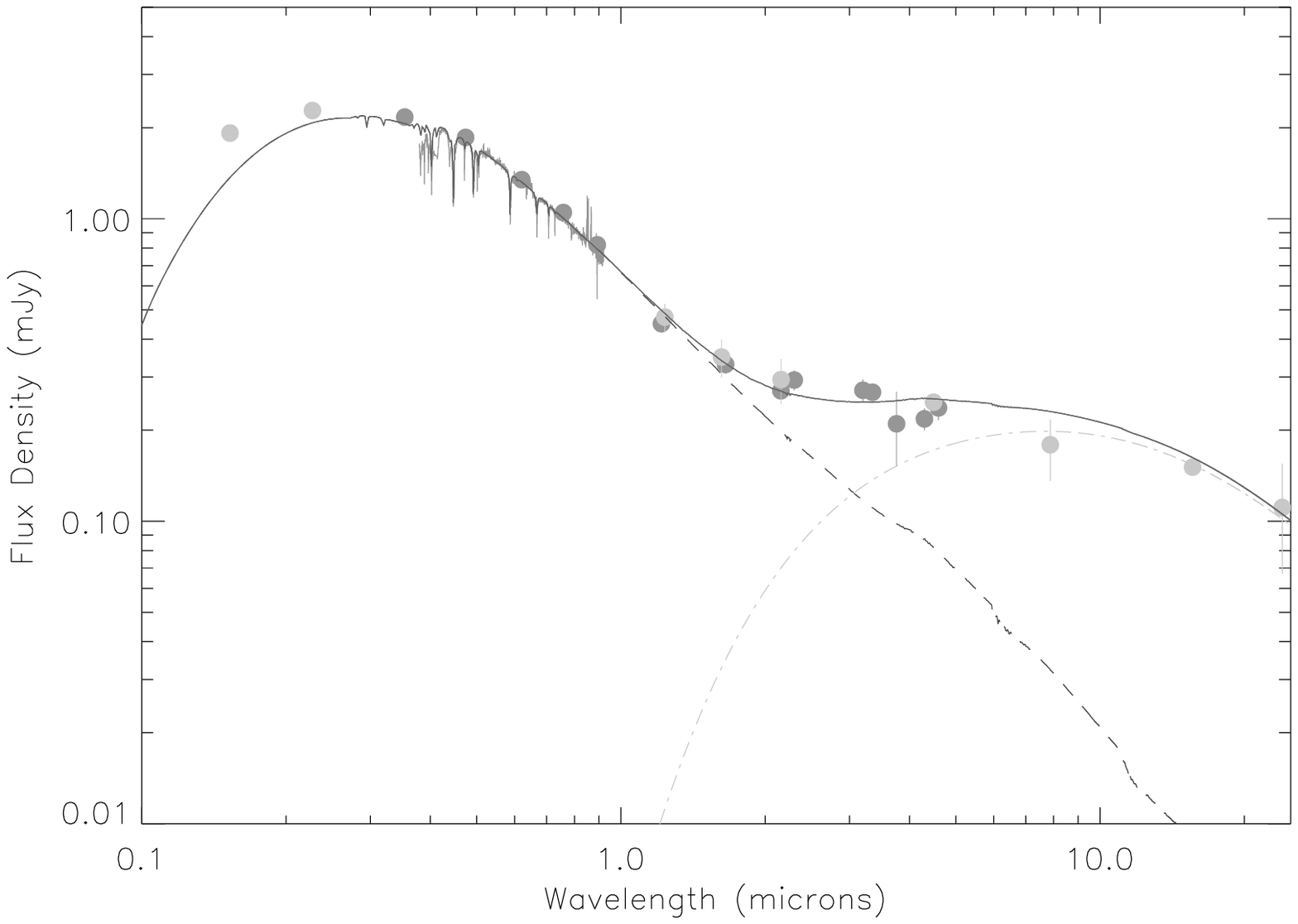}{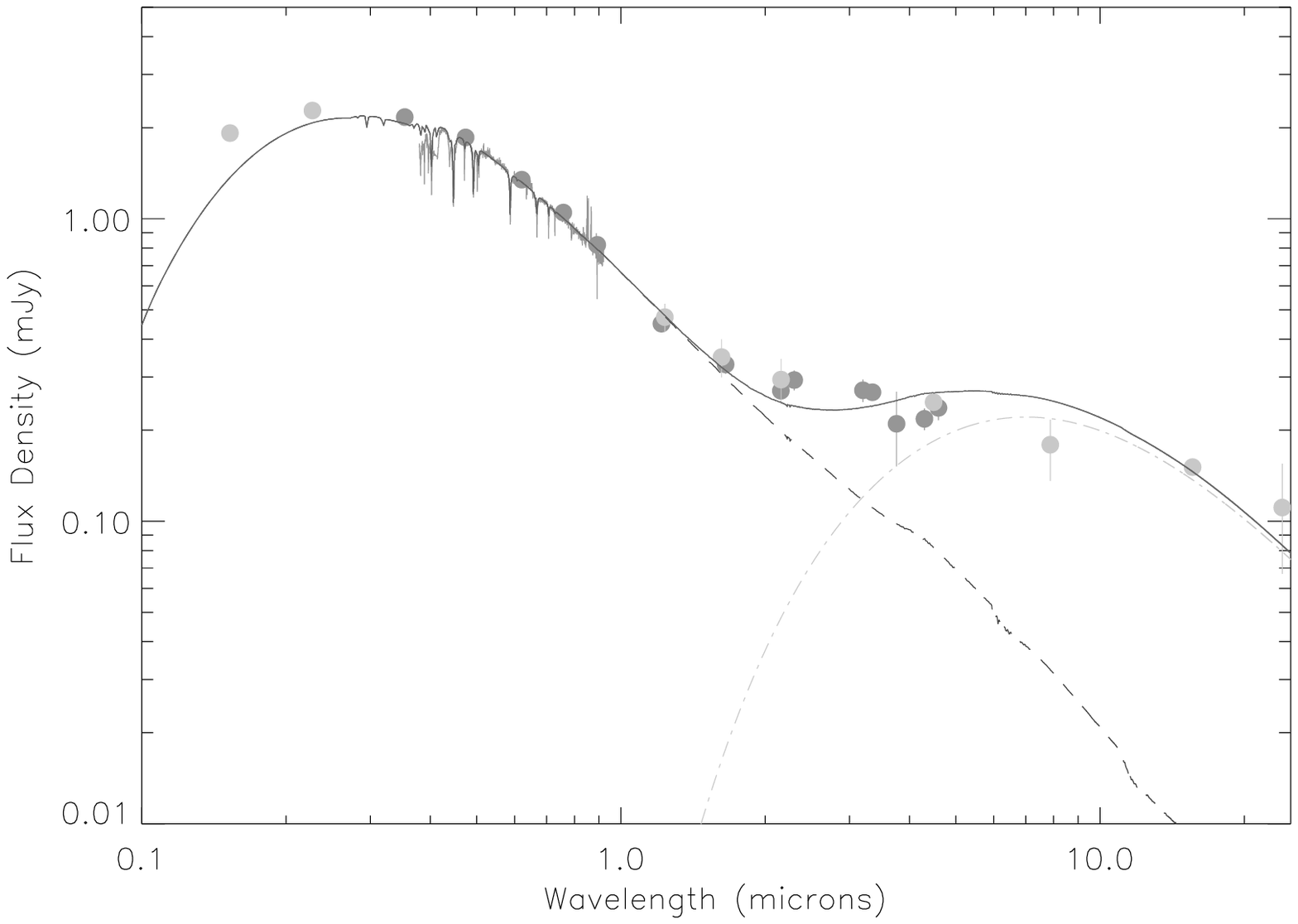}
\caption{The SED for SDSS\,J0845+2257 from the ultraviolet to the mid-infrared. GALEX, AAT and Spitzer data are plotted in light grey, while SDSS, NIR \citep{melis10}, AKARI \citep{farihi10} and WISE Preliminary Data Release photometry are plotted in dark grey, along with the SDSS DR7 white dwarf spectrum. The white dwarf model plus background galaxy (8 Gyr, z=1.6, \citet{maraston98,maraston05}; see Section\,\ref{section:sdss0845contaminant}) is plotted in dash-dot-dot-dot. The SED shows a significant flux density excess from the K-band to longer wavelengths, over the flux density expected from the white dwarf and the background galaxy alone. The SED on the left is compared with the best model of  a WD + background galaxy + optically thick dust disk with an inner dust temperature of 1800\,K and an outer disk temperature of 250\,K. The disk inclination is 83 degrees. We note that, while this is a good fit, the inner dust temperature is above the expected dust sublimation temperature.  The SED on the right is compared to the same model but with a physically-possible inner dust disk temperature of 1400\,K, an outer disk temperature of 400\,K and an inclination of 80 degrees. See Section\,\ref{section:sdss0845} for the full range of possible model parameters.}
\label{figure:sdss0845}
\end{figure}

The GALEX, AAT and Spitzer photometry for SDSS0845 are given in Table\,\ref{table:photometry}, and are plotted in light grey in Figure\,\ref{figure:sdss0845}, along with the SDSS DR7 spectrum, SDSS photometry, NIR photometry from \citet{melis10} and \citet{farihi10}, and the WISE Preliminary Data Release (dark grey). The data show a clear excess at wavelengths longer than 2\,$\mu$m, above the flux density expected from the white dwarf and the contaminating background galaxy. For the white dwarf contribution, we adopted a DB model with T$_{eff}$=18621\,K, log(g) = 8.2, d = 85\,pc, $R_{WD}$ = 0.009\,R$_{\odot}$. The resultant model is plotted as a dashed grey line, while the WD model with the background galaxy added is shown as a dash-dot-dot-dot line. 

The models for SDSS0845 are shown in Figure\,\ref{figure:sdss0845}. The first shows the best-fitting disk model, with an inner temperature of 1800\,K, an outer temperature of 250\,K, and an inclination of 85 degrees. Figure\,\ref{figure:sdss0845}b shows the closest physical model with an inner disk temperature constrained to 1400\,K, T$_{out} = 400\,K$ and inclination = 80 degrees.  This corresponds with inner and outer disk radii of $R_{in}$ = 18.8 $R_{WD}$ (0.17\,$R_{\odot}$) and $R_{out}$ = 99.7 $R_{WD}$ (0.89\,$R_{\odot}$). The coolest disk model still within $3\sigma$ of the plotted datapoints has $T_{in}$ = 1100\,K, $T_{out}$ = 600\,K and an inclination of 70 degrees. 

The outer radius of the gaseous component was estimated by Gaensicke et al. (2008) to be $\sim0.8-0.9R_{\odot}$, which is comparable to that of the dust disk derived above.

\subsection{SDSS\,J1043+0855}\label{section:sdss1043}

\begin{figure}
\plottwo{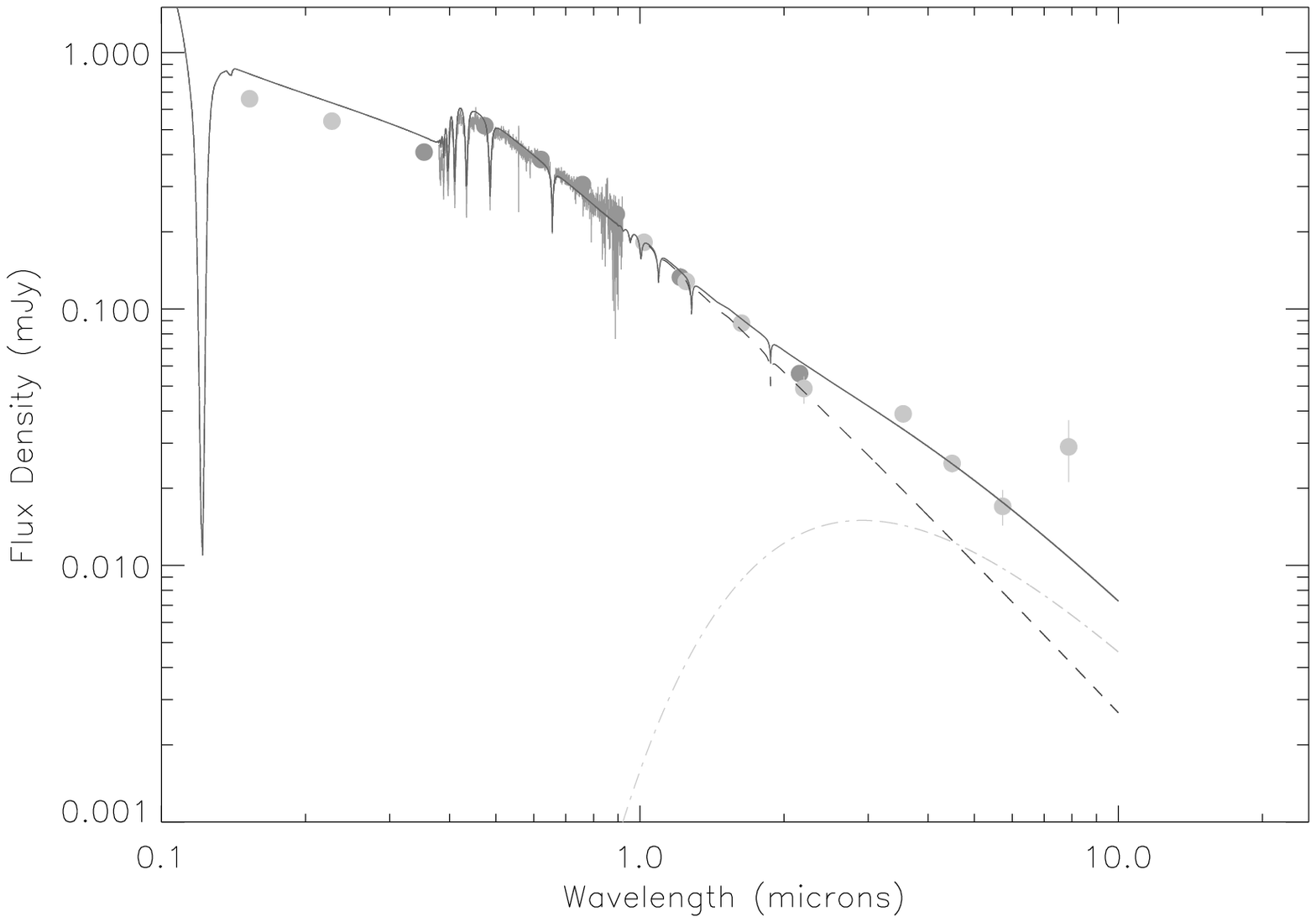}{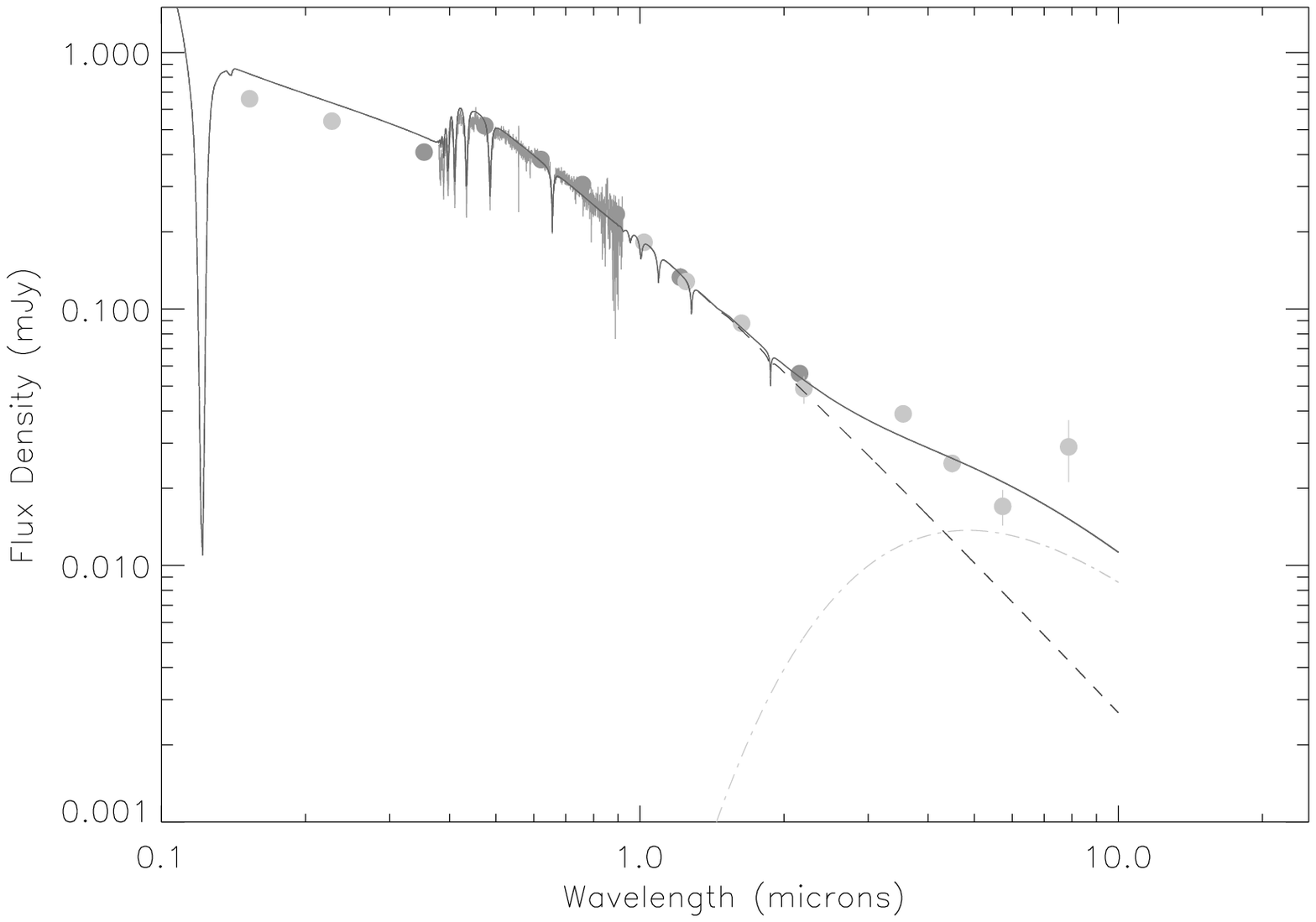}
\caption{The SED of SDSS\,1043+0855 from the ultraviolet to the mid-infrared. The WD spectrum was taken from SDSS Data Release 7 and is plotted along with the GALEX, UKIDSS and \textit{Spitzer} in light grey, and the SDSS photometry and NIR data from \citet{melis10} in dark grey. The SED shows a significant flux density excess above that expected from the white dwarf alone. The SED on the left is compared to a white dwarf + optically thick disk model with an inner dust temperature of 1800\,K, and outer disk temperature of 1700\,K and an inclination of 40 degrees. The SED on the right limits the inner disk temperature to $T_{in}$ = 1400\,K, with $T_{out}$ = 800\,K and an inclination of 85 degrees. See Section\,\ref{section:sdss1043} for the full range of possible models. }
\label{figure:sdss1043}
\end{figure}

The GALEX, UKIDSS and Spitzer flux densities for SDSS1043 are given in Table\,\ref{table:photometry}, and are plotted in Figure\,\ref{figure:sdss1043} together with the SDSS DR7 spectrum. The uncertainties on the Spitzer photometry for SDSS1043 are notably large, due to the faint nature of the source (only 30\,$\mu$Jy in IRAC channel 1). In addition to these data points, we include CFHT and Lick near-IR photometry from \citet{melis10} to constrain our model. All previously-published data are shown in dark grey in Figure\,\ref{figure:sdss1043}, while the GALEX, UKIDSS and Spitzer data are shown in light grey. We have fitted the DR7 spectrum of SDSS1043 using DA model spectra computed with the code of \citet{koester10}, which incorporates the updated Stark broadening calculations of \citet{tremblay09}, and find $T_{eff} = 17912\pm362\,K$ and $log(g) = 8.07\pm0.08$, corresponding to $M_{WD} = 0.66\pm0.05\,M_{\odot}$ and $R_{WD} = 8.67\pm0.47 \times 10^8$\,cm (Table 3). These values are, within the uncertainties, consistent with those derived from the DR4 spectrum by \citet{gaensicke07}, but given the improvement in the SDSS data reduction and line broadening, the new values reported here should be preferred. 

 The IRAC data points show a 3$\sigma$ excess above the expected flux density of the white dwarf, increasing dramatically at 8 microns. 

As with the other targets, we show the best-fit optically thick dust disk model in Figure\,\ref{figure:sdss1043}a, with an inner disk dust temperature of $T_{in}$ = 1800K, T$_{out}$ = 1700\,K and an inclination of 40 degrees. When constraining the inner disk temperature below the sublimation threshold of 1400\,K, we find a best-fit model with $T_{in}$ = 1400, $T_{out}$ = 800\,K and $i$ = 85 degrees.  This corresponds with inner and outer disk radii of $R_{in}$ = 17.9 $R_{WD}$ (0.23\,$R_{\odot}$)and $R_{out}$ = 37.6 $R_{WD}$ (0.49$R_{\odot}$). The coolest model allowed by the fit has $T_{in}$ = 900\,K, $T_{out}$ = 800\,K and $i$ = 56 degrees. 

Modelling the double-peaked CaII line profiles, \citet{gaensicke12} find, similar as for SDSS0738 and SDSS0845, $R_{max}\sim1R_{\odot}$ \citep[see also][]{melis10}.

\subsection{SDSS\,J1617+1620}\label{section:sdss1617}

\begin{figure}
\plottwo{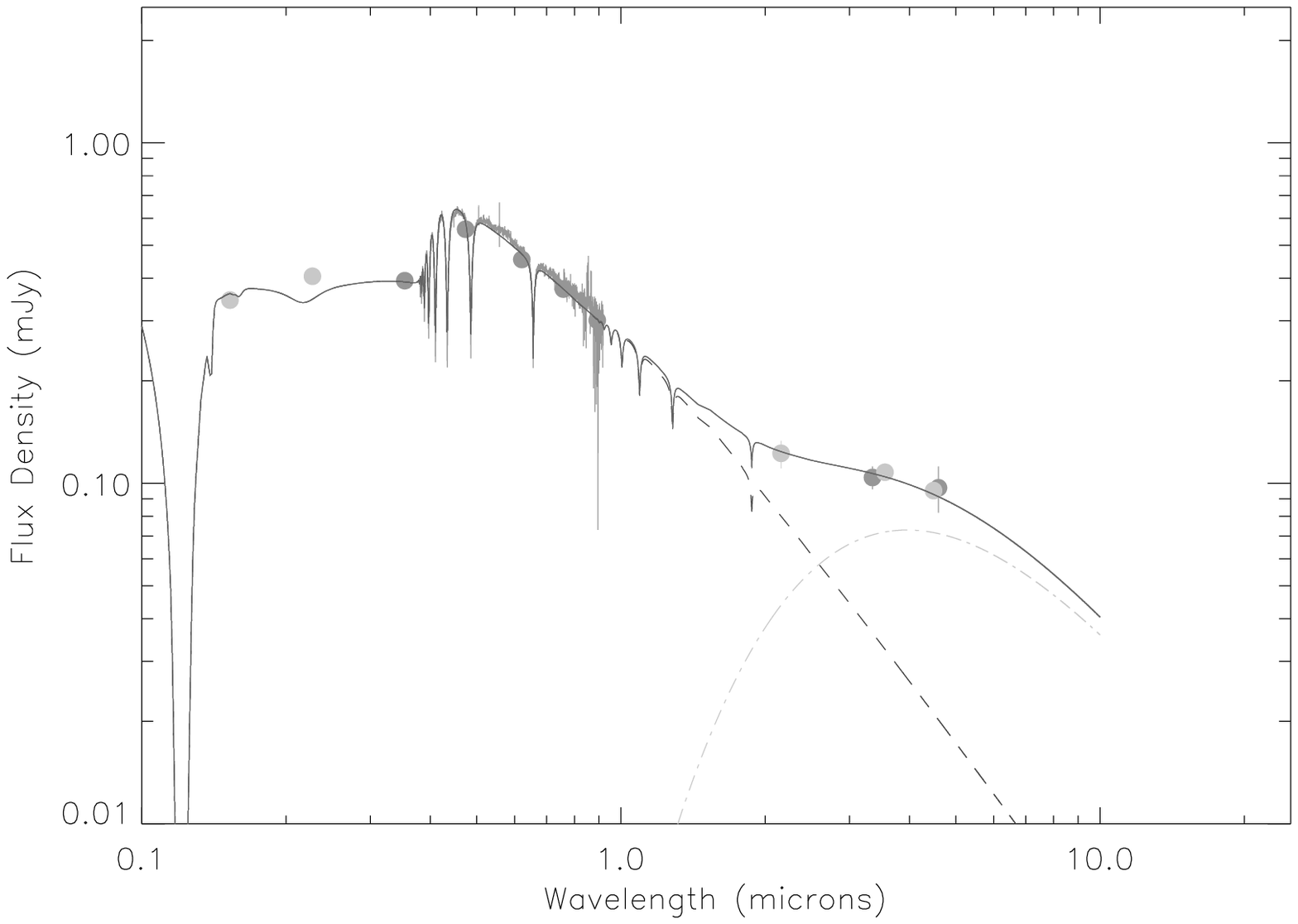}{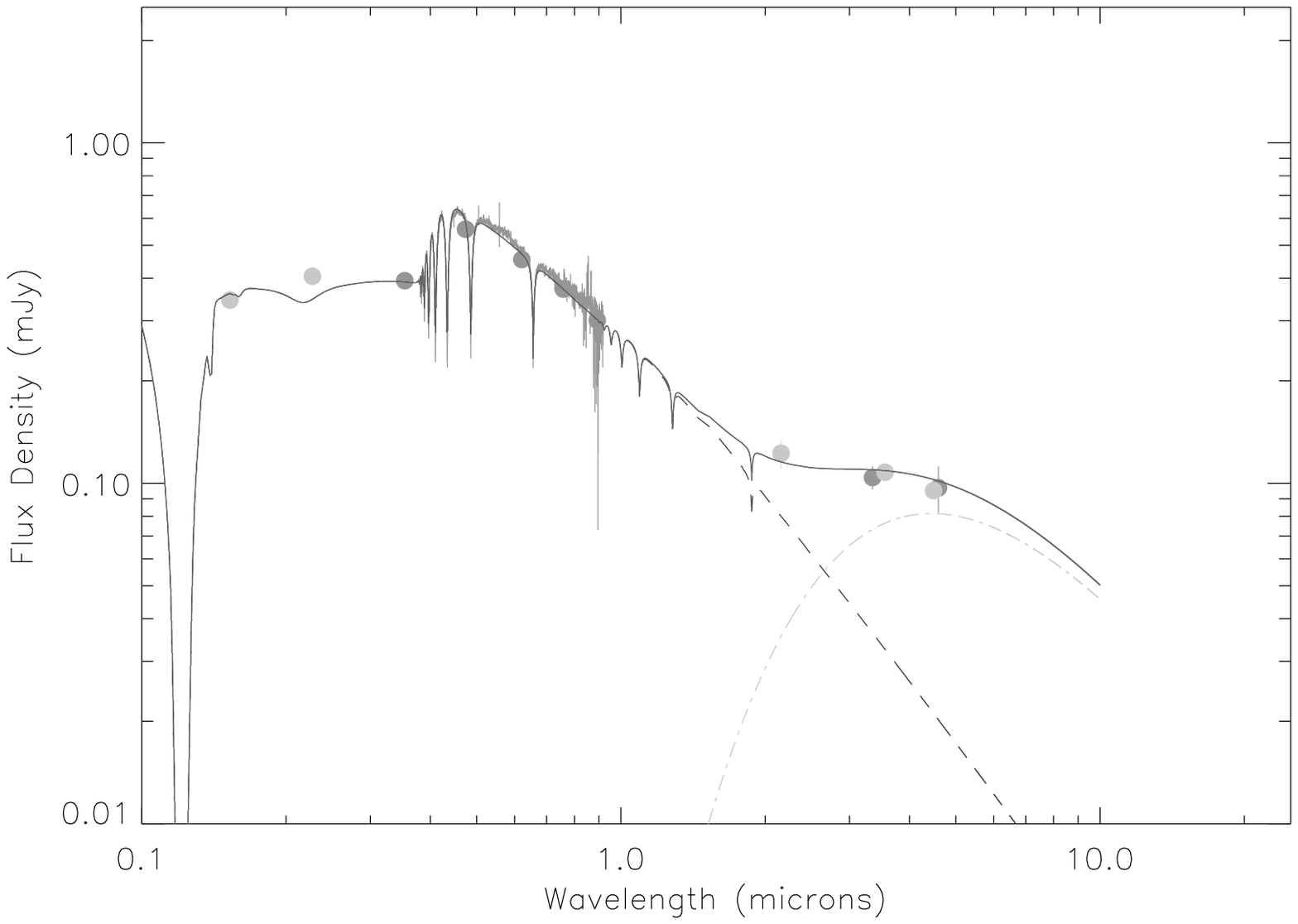}
\caption{The SED for SDSS\,J1617+1620 from the ultraviolet to the mid-infrared. \textit{Spitzer}, GALEX, WISE and SOFI data are reported in Table\,\ref{table:photometry}. GALEX, SOFI and Spitzer data are plotted in light grey, while the SDSS data and the WISE Preliminary Data Release are shown in dark grey. The WD spectrum was taken from SDSS Data Release 7. The SED shows a significant flux density excess from the K-band to longer wavelengths, over the flux density expected from the white dwarf alone. The SED on the left is compared to the best-fit white dwarf + optically thick disk model with an inner dust temperature of 1800\,K, and outer disk temperature of 950\,K and an inclination of 70 degrees. The SED on the right limits $T_{in}$ to 1400\,K with $T_{out}$ = 950\,K and $i$ = 50 degrees. See Section\,\ref{section:sdss1617} for the full range of model parameters.}
\label{figure:sdss1617}
\end{figure}

The flux densities from GALEX, SOFI and Spitzer are given in Table\,\ref{table:photometry}, and are plotted in light grey in Figure\,\ref{figure:sdss1617}, together with the SDSS DR7 spectrum (dark grey). The data show a clear excess above the flux density from the white dwarf at the Spitzer wavelengths. The white dwarf model, also based on \citet{gaensicke11}, is shown as a dashed black line and is a DA WD model, with  T$_{eff}$ = 13432\,K, log(g) = 8.04, d = 122\,pc, $R_{WD}$ = 0.01261\,R$_{\odot}$, reddened by E(B-V) = 0.04 to better match the optical SDSS spectrum. The reddening has negligible effect at infrared wavelengths. 

The best-fit dust disk model has an inner disk effective temperature T$_{in}$ = 1800\,K, outer disk $T_{out}$ = 950\,K and an inclination of 70 degrees, plotted as a grey dash-dot line in Figure\,\ref{figure:sdss1617}a. The closest fit with $T_{in}$ fixed at 1400K has $T_{out}$ = 950\,K and $i$ = 50 degrees (Figure\,\ref{figure:sdss1617}b).  This corresponds with inner and outer disk radii of $R_{in}$ = 12.2 $R_{WD}$ (0.15$R_{\odot}$) and $R_{out}$ = 20.4 $R_{WD}$ (0.26\,$R_{\odot}$). The model with the coolest possible inner disk temperature has $T_{in}$ = 1100\,K, $T_{out}$ = 620\,K and $i$ = 53 degrees. 

The peak separation of the CaII emission lines implies a maximum outer radius of the gaseous component of ~90$R_{WD}$ \citep{gaensicke12}.

\section{Discussion and Conclusions} 

We find that out of these four white dwarfs with previously-identified gaseous disks, all show evidence for dusty extensions to those disks, as seen in the previously-studied SDSS\,J1228+1040. We note that the best-fitting models for all of these disks have inner disk temperatures hotter than the sublimation temperature expected from the bulk composition of the dust, although all can be modelled satisfactorily with inner disk temperatures fixed to a physically more plausible temperature of 1400\,K.  When adding in the results from our previous study of SDSS1228 \citep{brinkworth09}, we find that all five white dwarfs with CaII emission from a gaseous debris disk also show dusty components to those disks. It is evident that gaseous and dusty debris disks easily coexist around white dwarfs, and the dusty disks likely feed the gaseous component. Both \citet{jura08} and \citet{melis10} suggest that dust disks are a pre-requisite to gaseous disks, and our findings are consistent with that evolutionary path, and vice-versa, the presence of gas is likely to have a major effect on the evolution of the  planetary debris discs around white dwarfs \citep{rafikov11b}.

The one outlier of our sample is SDSS1043. Even taking the uncertainties on the photometry into account, it is apparent that the dust in the system is far less pronounced than in the other three targets, despite very similar gaseous disk signatures in all four systems. The 4.5$\mu$m excesses seen in SDSS0738,  SDSS0845 and SDSS1617 are 7.5, 5.7, and 4.5 times the white dwarf contribution, respectively. In contrast, in SDSS1043, the measured flux density at 4.5 microns is only 1.7 times the white dwarf contribution. Why SDSS1043 should lack significant dust compared to the other three systems remains a mystery. Maybe the system is at a higher inclination, so we are seeing the correspondingly smaller projected area of the edge of the geometrically thin disk, rather than its face, but this doesn't seem to be borne out in the modelling of the other systems. Considering the likelihood that the dust disks feed the gaseous disks in these systems, it is feasible that we are seeing the tail-end of an asteroidal tidal disruption event that took place at an earlier time than the other systems, such that the dust around SDSS1043 is now depleted compared to the other three white dwarfs, and the gaseous disk dominates the SED. This seems consistent with the most recent estimates for disk lifetimes: a massive ($\sim10^{22}$) disk accreting due to Poynting-Robertson drag alone can be expected to survive for several Myr \citep{rafikov11a}. However, when a gas disk is present in the system due to sublimation of the inner disk, runaway accretion can occur. \citet{rafikov11b} predict that this can lead to the disk being exhausted in $\sim10^{5}$\,yr. \citet{bochkarev11} extend this analysis to include gas generated from multiple asteroid impacts, in which they confirm the estimated disk lifetime of several Myr. On the other hand, \citet{debes11} calculates that dust disks around white dwarfs may last for up to 1\,Gyr, while gas disks that are not being replenished are unlikely to last more than 10$^{4}$ years before fully accreting onto the white dwarf. If the disk around SDSS1043 is indeed reaching the end of its lifetime, long-term infrared monitoring of the SEDs of all of the five systems (SDSS1228, SDSS0738, SDSS0845, SDSS1043 and SDSS1617) over the next few decades may help to shed some light on this.

\acknowledgments

The authors would like to thank the referee for detailed and useful comments that helped to improve this paper. 

The DA white dwarf models used in the analysis were calculated with the modified Stark broadening profiles of \citet{tremblay09}, kindly made available by the authors. 

Co-authors Gaensicke and Marsh were supported by an STFC rolling grant. 

This work is based on observations made with the Spitzer Space Telescope, which is operated by the Jet Propulsion Laboratory, Caltech, under NASA contracts 1407 and 960785. This work makes use of data products from the Two Micron All Sky Survey, which is a joint project of the University of Massachusetts and IPAC/Caltech, funded by NASA and the NSF. The SOFI results in this paper are based on observations collected at the European Southern Observatory (La Silla) under programme ID 085.D-0541. This publication makes use of data products from the Wide-field Infrared Survey Explorer, which is a joint project of the University of California, Los Angeles, and the Jet Propulsion Laboratory/California Institute of Technology, funded by the National Aeronautics and Space Administration. Funding for the SDSS and SDSS-II has been provided by the Alfred P. Sloan Foundation, the Participating Institutions, the National Science Foundation, the U. S. Department of Energy, the National Aeronautics and Space Administration, the Japanese Monbukagakusho, the Max Planck Society and the Higher Education Funding Council for England. The SDSS Web Site is http://www.sdss.org/. The SDSS is managed by a very long list of institutions, who can all be found at the following website: http://www.sdss.org/collaboration/credits.html.

This research has made use of NASA's  Astrophysics Data System. This research has made use of the SIMBAD database, operated at CDS, Strasbourg, France.


\begin{deluxetable}{lccccc}
\tablecaption{Observation log for Spitzer Space Telescope data}
\tabletypesize{\scriptsize}
\tablehead{
\colhead{Target} & \colhead{Instrument/} & \colhead{AOR Key} & \colhead{Total Integration} & \colhead{Date} & \colhead{Pipeline} \\
\colhead{} & \colhead{Channel} & \colhead{} & \colhead{Time (s)} & \colhead{} & \colhead{} \\
}
\tablecolumns{6}
\startdata
SDSS\,J0845+2257 & IRAC Ch. 2   & 25459968 & 600  & 2008 May 09 & S18.7 \\
SDSS\,J0845+2257 & IRAC Ch. 4   & 25459968 & 600  & 2008 May 09 & S18.7 \\
SDSS\,J0845+2257 & IRS Blue PUI & 25460224 & 600  & 2008 Dec 02 & S18.7 \\
SDSS\,J0845+2257 & MIPS-24      & 25460480 & 140  & 2008 May 18 & S18.12 \\
SDSS\,J0845+2257 & MIPS-24      & 25460736 & 140  & 2008 May 19 & S18.12 \\
SDSS\,J1043+0855 & IRAC Ch. 1   & 22248448 & 3000 & 2008 Jun 10 & S18.7 \\
SDSS\,J1043+0855 & IRAC Ch. 2   & 22248448 & 3000 & 2008 Jun 10 & S18.7 \\
SDSS\,J1043+0855 & IRAC Ch. 3   & 22248448 & 3000 & 2008 Jun 10 & S18.7 \\
SDSS\,J1043+0855 & IRAC Ch. 4   & 22248448 & 3000 & 2008 Jun 10 & S18.7 \\
SDSS\,J1617+1620 & IRAC Ch. 1   & 39873024 & 750  & 2010 Sep 01 & S18.18 \\
SDSS\,J1617+1620 & IRAC Ch. 2   & 39873024 & 750  & 2010 Sep 01 & S18.18 \\
SDSS\,J0738+1835 & IRAC Ch. 1   & 39872768 & 3000 & 2010 Dec 01 & S18.18 \\
SDSS\,J0738+1835 & IRAC Ch. 2   & 39872768 & 3000 & 2010 Dec 01 & S18.18 \\
\enddata 
\label{table:observations}
\end{deluxetable}

\clearpage
\begin{deluxetable}{lccccc}
\rotate
\tablewidth{0pt}
\setlength{\tabcolsep}{0.05in}
\tablecaption{Photometry for all four target stars. }
\tablehead{
\colhead{Waveband} & \colhead{Wavelength ($\mu$m)} & \multicolumn{4}{c}{Flux Density (mJy)}\\
\colhead{} & \colhead{} & \colhead{SDSS0738+1835}     & \colhead{SDSS0845+2257}     & \colhead{SDSS1043+0855}         & \colhead{SDSS1617+1620}    \\
}
\tablecolumns{6}
\startdata
GALEX FUV & 0.158       & 0.061$^{+11\%}_{-12\%}$     & 1.92$\pm5\%$                & 0.66$\pm5\%$                    & 0.345$^{+4.2\%}_{-4.4\%}$       \\
GALEX NUV & 0.227       & 0.161$^{+4.7\%}_{-4.9\%}$   & 2.28$\pm5\%$                & 0.54$\pm5\%$                    & 0.406$^{+1.7\%}_{-1.8\%}$       \\
Y         & 1.02        & --                          & --                          & 0.182$^{+4.9\%}_{-4.4\%} (b)$ & --                              \\
J         & 1.2         & --                          & 0.473$\pm11\% (a)$        & 0.128$^{+6.3\%}_{-5.4\%} (b)$ & --                              \\
H         & 1.6         & --                          & 0.349$\pm14\% (a)$        & 0.088$^{+7.4\%}_{-6.9\%} (b)$ & --                              \\
K         & 2.2         & --                          & 0.294$\pm17\% (a)$        & 0.049$^{+13\%}_{-12\%} (b)$   & 0.122$^{+8.8\%}_{-9.6\%} (c)$ \\
WISE-1    & 3.35        & --                          & 0.267$\pm4.5\%$             & --                              & 0.104$\pm7.7\%$                 \\
IRAC-1    & 3.550       & 0.082$\pm3\%$               & --                          & 0.039$\pm6\%$                   & 0.108$\pm3\%$                   \\
IRAC-2    & 4.493       & 0.075$\pm3.7\%$             & 0.248$^{+3\%}_{-4\%}$       & 0.025$\pm8\%$                   & 0.095$\pm3\%$                   \\
WISE-2    & 4.60        & --                          & 0.237$\pm8.9\%$             & --                              & 0.097$\pm15.5\%$                \\
IRAC-3    & 5.731       & --                          & --                          & 0.017$\pm16\%$                  & --                              \\
IRAC-4    & 7.872       & --                          & 0.179$^{+21\%}_{-24\%}$     & 0.029$\pm27\%$                  & --                              \\
IRS PU-16 & 15.60       & --                          & 0.151$^{+3.6\%}_{-4.7\%}$   & --                              & --                              \\
MIPS-24   & 23.68       & --                          & 0.111$^{+4\%}_{-4\%}$       & --                              & --                              \\
\enddata 
\label{table:photometry}
\tablecomments{Source of NIR photometry: (a) AAT (b) UKIDSS (c) SOFI} 
\end{deluxetable}
\clearpage

\begin{deluxetable}{lcccc}
\tablecaption{Stellar parameters for target stars}
\tabletypesize{\scriptsize}
\tablehead{
\colhead{Parameter} & \colhead{SDSS\,J0738+1835}& \colhead{SDSS\,J0845+2257} & \colhead{SDSS\,J1043+0855} & \colhead{SDSS\,J1617+1620}  \\ 
}
\tablecolumns{5}
\startdata
WD Temperature (K) & 13600  & 18600  & 17912  & 13432   \\
Distance (pc)      & 120    & 85     & 183    & 122     \\
log(g)             & 8.5    & 8.2    & 8.0    & 8.0     \\
Radius (cm)        & 6.16e8 & 6.27e8 & 9.05e8 & 8.77e8  \\
\enddata
\label{table:parameters}
\end{deluxetable}


\begin{thebibliography}{99}

\bibitem[Abazajian et al.(2009)]{abazajian09} Abazajian et al. 2009, ApJ, 182, 543

\bibitem[Adelman-McCarthy et al.(2006)]{adelman-mccarthy06} Adelman-McCarthy, J.\,K., Ag\"ueros, M.\,A., Allam, S.\,S., Anderson, K.\,S.\,J., Anderson, S.\,F., Annis, J., Bahcall, N.\,A., Baldry, I.\,K., Barentine, J.\,C., Berlind, A. and 131 coauthors, 2006, ApJS, 162, 38

\bibitem[Becklin et al.(2005)]{becklin05} Becklin, E.~E., Farihi, J., Jura, M., Song, I., Weinberger, A.~J., \& Zuckerman, B.\ 2005, \apjl, 632, L119

\bibitem[Bertin \& Arnouts(1996)]{bertin96} Bertin, E., Arnouts, S., 1996, A\&AS, 117, 393

\bibitem[Bochkarev \& Rafikov(2011)]{bochkarev11} Bochkarev, K.~V., Rafikov, R. 2011, ApJ, 741, 36

\bibitem[Brinkworth et al.(2009)]{brinkworth09} Brinkworth, C. S., G\"ansicke, B. T., Marsh, T. R., Hoard, D. W., Tappert, C. \ 2009, ApJ, 696, 1402

\bibitem[Brinkworth et al.(2007)]{brinkworth07} Brinkworth, C.\,S., Hoard, D.\,W., Wachter, S., Howell, S.\,B., Ciardi, D.\,R., Szkody, P., Harrison, T.\,E., van Belle, G.\,T., Esin, A.\,A. 2007, ApJ, 659, 1541

\bibitem[Debes et al.(2011a)]{debes11a} Debes, J.~H., Hoard, D.~W., Kilic, M., Wachter, S., Leisawitz, D.~T., Cohen, M., Kirkpatrick, J.~D., Griffith, R.~L. 2011, ApJ, 729, 4

\bibitem[Debes(2011b)]{debes11} Debes, J. 2011b, White Dwarf Atmospheres and Circumstellar Environments (Ed. D. W. Hoard), Wiley.  

\bibitem[Debes \& Sigurdsson(2002)]{debes02} Debes, J.\,H., Sigurdsson, S. 2002, ApJ, 572, 556

\bibitem[Dufour et al.(2010)]{dufour10} Dufour, P., Kilic, M., Fontaine, G., Bergeron, P., Lachapelle, F, -R., Kleinman, S.  J., Leggett, S. K., 2010, ApJ, 719, 803

\bibitem[Dufour et al.(2012)]{dufour12} Dufour, P., Kilic, M., Fontaine, G., Bergeron, P., Melis, C., Bochanski, J., 2012, ApJ, accepted. arXiv:1201.6252

\bibitem[Fazio et al.(2005)]{fazio05} Fazio, G., et al.\ 2005, \apjs, 154, 10

\bibitem[Farihi et al.(2005)]{farihi05} Farihi, J., Becklin, E. E., Zuckerman, B., 2005, ApJS, 161, 394

\bibitem[Farihi et al.(2008)]{farihi08} Farihi, J., Zuckerman, B., Becklin, E.\,E., 2008, ApJ, 674, 431 

\bibitem[Farihi et al.(2009)]{farihi09} Farihi, J., Jura, M., Zuckerman, B. \ 2009, ApJ, 694, 805

\bibitem[Farihi et al.(2010)]{farihi10} Farihi, J., Jura, M., Lee, J.-E., Zuckerman, B. \ 2010, ApJ, 714, 1386

\bibitem[Ferguson et al.(2004)]{ferguson04} Ferguson, H.\,C., Dickinson, M., Giavalisco, M., Kretchmer, C., Raavindranath, S., Idzi, R., Taypor, E., Conselice, C.\,J., Fall, S.\,M., Gardner, J.~P., Livion, M., Madau, P., Moustakas, L,~A., Papovich, C.~M., Somerville, R.~S., Spinrad, H., Stern, D. 2004, ApJL, 600, L107

\bibitem[G\"ansicke et al.(2006)]{gaensicke06} G\"ansicke, B.\,T., Marsh, T.\,R., Southworth, J., Rebassa-Mansergas, A. \ 2006, Science, 314, 1908

\bibitem[G\"ansicke et al.(2007)]{gaensicke07} G\"ansicke, B.\,T., Marsh, T.\,R., Southworth, J. \ 2007, MNRAS, 380, L35

\bibitem[G\"ansicke et al.(2008)]{gaensicke08} G\"ansicke, B. T., Koester, D., Marsh. T. R., Rebassa-Mansergas, A., Southworth, J. \ 2008, MNRAS, 391L, 103

\bibitem[G\"ansicke et al.(2011)]{gaensicke11} G\"ansicke, B. T. 2011, in \textit{Planetary systems beyond the main sequence}, AIP Conf. Proc. 1331, 211(Eds. S. Schuh, H.~Drechsel,\& U.~Heber)

\bibitem[G\"ansicke et al.(2012)]{gaensicke12} G\"ansicke, B. T., et al. (2012) in prep.

\bibitem[Girven et al.(2012)]{girven11} Girven, J., Brinkworth, C.~S., Farihi, J., G\"ansicke, B.~T., Hoard, D.~W., Marsh, T.~R., Koester, D. 2012, accepted ApJ. arXiv:1202.3784v1

\bibitem[Graham et al.(1990)]{graham90} Graham, J.\,R., Matthews, K., Neugebauer, G., Soifer, B.\,T., 1990, ApJ, 357, 216

\bibitem[Hambly et al.(2008)]{hambly08} Hambly, N.\,C., Collins, R.\,S., Cross, N.\,J.\,G., Mann, R.\,G., Read, M.\,A., Sutorius, E.\,T.\,W., Bond, I., Bryant, J., Emerson, J.\,P., Lawrence, A. and 7 co-authors. 008, MNRAS, 384, 637

\bibitem[Hewett et al.(2006)]{hewett06} Hewett, R.\,J., Warren, S.\,J., Leggett, S.\,K., Hodgkin, S.\,T. 2006, MNRAS, 367, 454

\bibitem[Hoard et al.(2007)]{hoard07} Hoard, D.\,W., Howell, S.\,B., Brinkworth, C.\,S., Ciardi, D.\,R., Wachter, S. 2007, ApJ, 671, 734

\bibitem[Houck et al.(2004)]{houck04} Houck, J. R., et al. \ 2004, ApJS, 154, 211

\bibitem[Jura(2003)]{jura03} Jura, M. \ 2003, ApJ, 584, L91

\bibitem[Jura et al.(2007)]{jura07} Jura, M., Farihi, J., Zuckerman, B. \ 2007, ApJ, 663, 1285

\bibitem[Jura(2008)]{jura08} Jura, M., AJ, 135, 1785

\bibitem[Kilic et al.(2005)]{kilic05} Kilic, M., von Hippel, T., Leggett, S.\,K., Winget, D.\,E. \ 2005, ApJ, 632, L115

\bibitem[Kilic et al.(2006)]{kilic06} Kilic, M., von Hippel, T., Leggett, S.\,K., Winget, D.\,E. \ 2006, ApJ, 646, 474

\bibitem[Kilic \& Redfield(2007)]{kilic07} Kilic, M., Redfield, S. \ 2006, ApJ, 660, 641

\bibitem[Kilic et al.(2011)]{kilic11} Kilic, M., Patterson, A.~J., Barber, S., Leggett, S.~K., Dufour, P. 2011, MNRAS, 419, L59

\bibitem[Klein et al.(2011)]{klein11} Klein, B., Jura, M., Koester, D., Zuckerman, B. 2011, ApJ, 741, 64K

\bibitem[Koester et al.(1997)]{koester97} Koester, D., Provencal, J., Shipman, H.\,L. \ 1997, A\&A, 320, L57

\bibitem[Koester(2008)]{koester08} Koester, D. \ 2008, arXiv0812.0482

\bibitem[Koester(2010)]{koester10}Koester, D. 2010, MmSAI, 81, 921

\bibitem[Kuchner et al.(1998)]{kuchner98} Kuchner, M.\,J., Koresko, C.\,D., Brown, M.\,E., 1998, ApJ, 508, L81

\bibitem[Lawrence et al.(2007)]{lawrence07} Lawrence, A., Warren, S.\,J., Almaini, O., Edge, A.\,C., Hambly, N.\,C, Jameson, R.\,F., Lucas, P., Casili, M., Adamson, A., Dye, S. and 12 co-authors. 2007, MNRAS, 379, 1599

\bibitem[Lodders(2003)]{lodders03} Lodders. K. 2003, ApJ, 591, 1220

\bibitem[Makovoz et al.(2006)]{makovoz06} Makovoz, D., Roby, T., Khan, I., Booth, H., 2006, SPIE, 6274, 10

\bibitem[Maraston, C.(2005)]{maraston05} Maraston, C. 2005, MNRAS, 362, 799

\bibitem[Maraston et al.(1998)]{maraston98} Maraston, C. 1998, MNRAS, 300, 872

\bibitem[Martin et al.(2005)]{martin05} Martin et al. 2005, ApJ, 619, 1

\bibitem[Melis et al.(2010)]{melis10} Melis, C., Jura, M., Albert, L., Klein, B., Zuckerman, B. \ 2010, ApJ, 722, 1078

\bibitem[Patten et al.(2006)]{patten06} Patten, B., et al.\ 2006, ApJ, 651, 502

\bibitem[Rafokov(2011a)]{rafikov11a} Rafikov, R. 2011, ApJL, 732, L3

\bibitem[Rafikov(2011b)]{rafikov11b} Rafikov, R. 2011, MNRAS, 416L, 55

\bibitem[Reach et al.(2005)]{reach05} Reach, W.\,T., Kuchner, M.\,J., von Hippel, T., Burrows, A., Mullally, F., Kilic, M., Winget, D.\,E. \ 2005, ApJ, 635, L161

\bibitem[Rieke et al.(2004)]{rieke04} Rieke, G. H. et al. \ 2004, ApJS, 154, 25

\bibitem[Rousselot et al.(2000)]{rousselot00} Rousselot, P., Lidman, C., Cuby, J.-G., Moreels, G., Monnet, G. 2000, A\&A, 354, 1134

\bibitem[Skrutskie et al.(2006)]{skrutskie06} Skrutskie et al. 2006, AJ, 131, 1163 

\bibitem[Telesco, Joy \& Sisk(1990)]{telesco90} Telesco, C. M., Joy, M., Sisk, C., 1990, ApJL, 358, 17

\bibitem[Tokunaga, Becklin \& Zuckerman(1990)]{tokunaga90} Tokunaga, A. T., Becklin, E. E., Zuckerman, B., 1990, ApJL, 358, 21

\bibitem[Tremblay \& Bergeron(2009)]{tremblay09} Tremblay, P.-E., Bergeron, P. 2009, ApJ, 696,1755

\bibitem[Vennes, Kawka \& N\'emeth(2011)]{vennes11} Vennes, S. Kawka, A., N\'emeth, P. 2011, MNRAS, 413, 2545

\bibitem[von Hippel et al.(2007)]{vonhippel07}von Hippel, T., Kuchner, M.\,J., Kilic, M., Mullally, F., Reach, W.\,T. \ 2007, ApJ, 662, 544

\bibitem[Xu \& Jura(2011)]{xu11} Xu, S., Jura, M., 2011, ApJ, in press. arXiv:1109.4207

\bibitem[Zuckerman \& Becklin(1987)]{zuckerman87} Zuckerman, B., \& Becklin, E.\,E. \ 1987, Nat., 330, 138

\bibitem[Zuckerman et al.(2007)]{zuckerman07} Zuckerman, B., Koester, D., Melis, C., Hansen, B.~M., Jura, M. 2007, ApJ, 671, 872  

\end{thebibliography}
\end{document}